\documentclass[english]{IEEEtran}
\usepackage[T1]{fontenc}
\usepackage[latin9]{inputenc}
\usepackage{babel}
\usepackage{bm}
\usepackage{amsmath}
\usepackage{amsthm}
\usepackage{amssymb}
\usepackage{graphicx}
\usepackage[unicode=true,
 bookmarks=true,bookmarksnumbered=true,bookmarksopen=true,bookmarksopenlevel=1,
 breaklinks=false,pdfborder={0 0 0},pdfborderstyle={},backref=false,colorlinks=false]
 {hyperref}
\hypersetup{pdftitle={Your Title},
 pdfauthor={Your Name},
 pdfpagelayout=OneColumn,pdfnewwindow=true,pdfstartview=XYZ}

\makeatletter

\let\SF@@footnote\footnote
\def\footnote{\ifx\protect\@typeset@protect
    \expandafter\SF@@footnote
  \else
    \expandafter\SF@gobble@opt
  \fi
}
\expandafter\def\csname SF@gobble@opt \endcsname{\@ifnextchar[
  \SF@gobble@twobracket
  \@gobble
}
\edef\SF@gobble@opt{\noexpand\protect
  \expandafter\noexpand\csname SF@gobble@opt \endcsname}
\def\SF@gobble@twobracket[#1]#2{}

\theoremstyle{plain}
\newtheorem{thm}{\protect\theoremname}
\theoremstyle{definition}
\newtheorem{defn}[thm]{\protect\definitionname}

\usepackage{babel}
\ifCLASSOPTIONcompsoc
\else
\fi
\usepackage{psfrag}
\usepackage{epsfig}\usepackage{graphics}\usepackage{cite}
\ifCLASSOPTIONcompsoc
\usepackage[caption=false,font=normalsize,labelfont=sf,textfont=sf]{subfig}
\else
\usepackage[caption=false,font=footnotesize]{subfig}
\fi

\@ifundefined{showcaptionsetup}{}{%
 \PassOptionsToPackage{caption=false}{subfig}}
\usepackage{subfig}
\makeatother

\providecommand{\definitionname}{Definition}
\providecommand{\theoremname}{Theorem}

\begin{document}

\title{Proactive Resource Allocation with Predictable Channel Statistics}

\author{L.~Srikar~Muppirisetty, John Tadrous,~\IEEEmembership{Member,~IEEE},
Atilla Eryilmaz,~\IEEEmembership{Member,~IEEE} and Henk~Wymeersch,~\IEEEmembership{Member,~IEEE}
\thanks{This research was supported, in part, by the European Research Council,
under Grant No. 258418 (COOPNET); by STINT Grant IB2013-5301; by NSF
Grants: CAREER-CNS-0953515, CNS-WiFiUS-1456806, and CCSS-EARS-1444026;
and the DTRA Grant: HDTRA1-15-1-0003. Also, the work of A. Eryilmaz
was supported, in part, by the QNRF Grant: NPRP 7-923-2-344. 

L. Srikar Muppirisetty was with Chalmers University of Technology
is now with the Volvo Car Corporation, Sweden. (E-mail: srikar.muppirisetty@volvocars.com).
Henk Wymeersch is with Chalmers University of Technology, Sweden (E-mail:
henkw@chalmers.se). John Tadrous is with Gonzaga University, USA (E-mail:tadrous@gonzaga.edu).
Atilla Eryilmaz is with The Ohio State University, USA (E-mail: eryilmaz.2@osu.edu).
\protect \\
Part of this work was presented in \cite{muppirisetty2015proactive}. }}
\maketitle
\begin{abstract}
The behavior of users in relatively predictable, both in terms of
the data they request and the wireless channels they observe. In this
paper, we consider the statistics of such predictable patterns of
the demand and channel jointly across multiple users, and develop
a novel predictive resource allocation method. This method is shown
to provide performance benefits over a reactive approach, which ignores
these patterns and instead aims to satisfy the instantaneous demands,
irrespective of cost to the system. In particular, we show that our
proposed method is able to attain a novel fundamental bound on the
achievable cost, as the service window grows. Through numerical evaluation,
we gain insights into how different uncertainty sources affect the
decisions and the cost.
\end{abstract}

\section{Introduction}

With the increase in number of users and data traffic per users, come
major challenges for network operators, leading to a need for more
intelligence at the network side \cite{phdthesisSrikar}. In the last
decade, the predictability of users has been assessed, indicating
that users are is not completely random in terms of their demand patterns
\cite{song2010limits,wang2011human,jensen2010estimating} or in the
wireless channel qualities they observe \cite{location2014rocco,spatial2015srikar,malmirchegini2012spatial,kim2011cooperative}.
This latter effect is a consequence of fixed mobility and behavioral
patterns, as well as the characteristics of wireless propagation.

Unlike conventional scheduling, which operates at the millisecond
time scale, predictability can be exploited for scheduling at a much
slower time scale (seconds, minutes) \cite{piro2011two,bui2016anticipatory}.
Networks currently assign resources to users at the slower time scale
in a reactive manner, regardless of channel quality (e.g., by setting
priority levels) \cite{capozzi2013downlink}. Future networks on the
other hand can collect data regarding users' demand and mobility patterns,
which can be combined with user predictability to forecast the user's
demand and channel. This leads to extra degrees of freedom to be exploited,
by allocating resources ahead of time. The main advantage of this
approach is network load balancing over large time scale dynamics,
at the expense of possible waste of network resources \cite{John2014}.

Predictive resource optimization schemes for efficient video content
streaming based on user channel quality metrics (CQM) are studied
in \cite{atawia2014robust,Abou2013Optimal,abou2013predictive,mekki2015anticipatory}.
Under known demands, \cite{lee2013asymptotically,Zafer2009} analyzed
energy-efficient policies for scheduling with statistical CQM knowledge.
 In \cite{location2014rocco,spatial2015srikar,malmirchegini2012spatial,kim2011cooperative},
a location-aided framework was proposed and showed how large-scale
channel characteristics of the wireless channel can be predicted by
exploiting the user's location information. Since location-aided predicted
CQM is coarse, it can be efficiently harnessed in predictive/proactive
resource allocation whereby demand dynamics and large-scale channel
characteristics vary within the same time scale. These above works
focus on channel predictability, ignoring the demand statistics. Demand
statistics were considered in the following works. In \cite{Minghua2014a,Minghua2014b},
user delay was evaluated under proactive scheduling and was found
to be reduced the longer the prediction window. The predictable demands
were exploited in \cite{tadrous2013proactive,John2014}: \cite{tadrous2013proactive}
proposed a proactive resource allocation framework, while \cite{John2014}
derived lower bounds on the cost of such proactive resource allocation
with time-varying user demands, as well as policies that can asymptotically
attain these bounds, again as the prediction window is increased.
The joint treatment of channel predictability and user demand predictability
is not considered in these works, which is the gap the current paper
aims to address.

In this work, we study proactive resource allocation strategies that
exploit both the predictable data demand and channel characteristics,
with uncertainties. Our main focus will be on time-varying, but predictable
channel statistics, as would be experienced when a user traverses
a known path. The main contributions of this paper can be summarized
as follows: 
\begin{itemize}
\item We extend the work in \cite{John2014,muppirisetty2015proactive}:
while \cite{John2014} did not consider the effect of the channel,
here it is included explicitly. Moreover, our preliminary study \cite{muppirisetty2015proactive}
focused on time-invariant channels, while we here explicitly consider
the effect of time-varying channel statistics. In addition, we allow
for general correlation among demands and among user statistics. 
\item We establish global lower bounds on the proactive scheduling cost
that capture the impact of demand and channel uncertainties. Specifically,
we compute lower bounds for two scenarios: (i) time-invariant demand
and channel statistics, (ii) time-invariant demand and time-varying
channel statistics.
\item We develop asymptotically optimal service policies that can attain
the bounds as the proactive service window grows in size. 
\item Through Monte Carlo simulations, we show the performance benefits
of the proactive schedulers over a reactive scheduler, in terms of
channel load and network cost. 
\end{itemize}
The remainder of the paper is structured as follows. Section \ref{sec:System-Model}
presents the system model comprising user demand model, wireless channel
model, and reactive and proactive network models. Section \ref{sec:Channel-Predictability}
provides empirical support for the key assumption related to the channel
for the proactive network model. In Section \ref{sec:Proactive-TITI},
we present a global lower bound for both time-invariant and time-varying
channel statistics, as well as asymptotically optimal stationary policies.
Finally, numerical results are given in Section \ref{sec:Numerical-Results},
followed by the conclusions in Section \ref{sec:Conclusions}. 

\subsubsection*{Notation}

Vectors and matrices are written in bold (e.g., a vector $\mathbf{k}$
and a matrix\textbf{ $\mathbf{K}$}); $\mathbb{E}[.]$ denotes expectation;
$[a_{n}]_{n=1}^{N}$ is a shorthand for $[a_{1},\ldots,a_{N}]^{T}$;
$\delta_{n\in\mathcal{B}}=1$ when $n\in\mathcal{B}$ and zero otherwise. 

\section{System Model\label{sec:System-Model}}

We consider a network system wherein the data requests from a set
of users are serviced by the SP. The SP provides service to each user
on per time slot basis considering the demand request from the users.
We assume that the network consists of set of $N$ users $\mathcal{N}=\{1,2,\ldots n,\ldots,N$\}.
The time slots are indexed by $t$ and all time slots all of them
have fixed duration. The users send data requests to the SP based
on their data requirements. The data request generation from the user
is capture by a binary random variable $d_{n,t}\in\{0,1\}$. When
$d_{n,t}=1$, the user $n$ generates a data request in time slot
$t$, and when $d_{n,t}=0$, there is no data request from user $n$.
Demands may be correlated among users (e.g., based on their social
connection). We consider a discrete random variable $g_{n,t}\in\mathbb{R}_{+}$
to capture the user experienced channel quality. The channel qualities
may also be correlated among users (e.g., if they are in close proximity).
The SP spends $S$ amount of resources for service each request from
the user.\footnote{The results obtained in this work can directly be generalized to the
case where such amount of resources is user and time-dependent, i.e.,
$S_{n,t}$, yet known to the system. } The statistics of the random variables of $d_{n,t}$ and $g_{n,t}$,
as well as the cost function to be minimized by SP are described below.
\begin{figure}
\centering

\includegraphics[width=1\columnwidth]{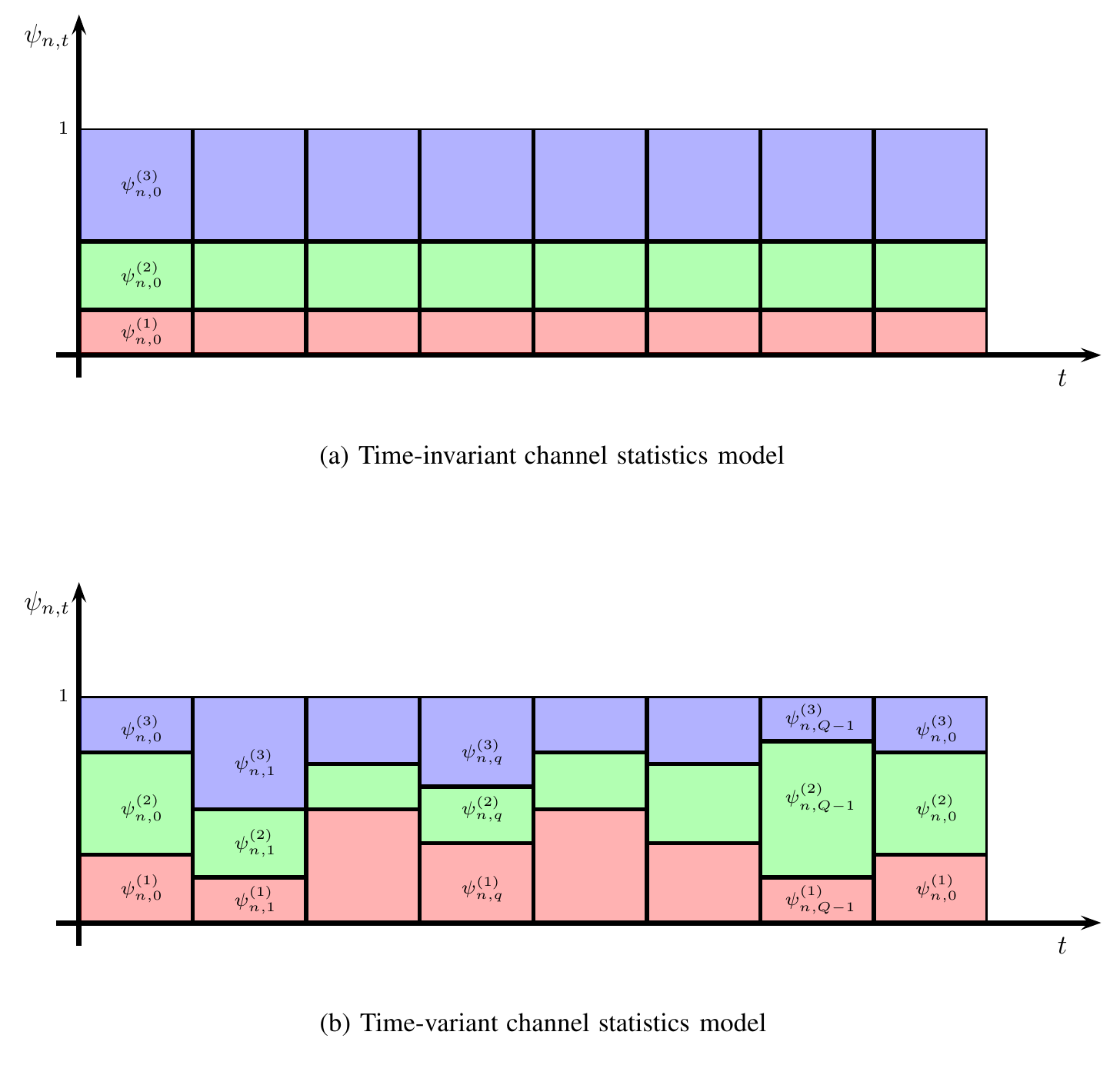}

\caption{\label{fig:Channel-statistics-model}Channel statistics model: Inset
(a) time-invariant channel statistics in which the channel state probabilities
do not vary with time. Inset (b) time-varying channel statistics,
wherein channels exhibit a cyclo-stationarity behavior with period
$Q$. }
\end{figure}

\subsection{Demand and channel statistics}

\subsubsection{User demand statistics}

We assume that the current data request of each user is known to the
SP. Furthermore, we consider a time-invariant demand statistics model,
where the demand probabilities do not vary in each $t$. The time-invariant
demand statistics model is of the form $\pi_{n,t}=p(d_{n,t}=1)$ for
all times $t$. For simplicity of exposition, we will assume the statistics
to be constant\footnote{The system can further be generalized to time-varying (fluctuating)
demand characteristics as in \cite{John2014}, yet this will lead
to complicated notations without significant conceptual benefit as
we focus on the impact of channel predictability. Hence, we have not
considered this scenario here. In addition, time-invariant demands
are reasonable for time scales on the order to tens of seconds or
minutes, during which channel statistics can change significantly. }, with $d_{n,t}$ being i.i.d.~with $\mathbb{E}[d_{n,t}]=\bar{\pi}_{n}$.
We can capture demand profile of all the users in a set $\bar{\boldsymbol{\pi}}=\{\bar{\pi}_{1},\ldots,\bar{\pi}_{N}\}$
and is known to the SP. User demand requests cannot be delayed but
can be serviced beforehand, which means the SP has to offer service
to a request in time $t$ no later than at time $t$.

\subsubsection{Channel statistics \label{subsec:Wireless-channel-model}}

The channel gain experienced by user $n$ in time slot $t$ is denoted
by $g_{n,t}$. As aforementioned, $g_{n,t}$ is modeled as discrete
random variable with $K_{n}$ states. The $K_{n}$ states are from
a finite set $\mathcal{C}_{n}=\{g_{n}^{(k)},\,k=1,\ldots,K_{n}$\}.
The statistics of $g_{n,t}$ are described by $\bm{\psi}_{n,t}=\{\psi_{n,t}^{(k)},k=1,\ldots,K_{n}\}$.
The channel is considered to be cyclo-stationary, so that the $\bm{\psi}_{n,t}$
is periodic in $t$ with period $Q$, which is assumed to be the same
for all users. Hence, the channel statistics of all users are determined
by 
\[
\boldsymbol{\Psi}^{Q}=\left\{ \mathcal{C}_{n},\bm{\psi}_{n,0},\ldots,\bm{\psi}_{n,Q-1}\right\} _{n=1}^{N}.
\]
As a special case, when $Q=1$, the channel becomes time-invariant
and is characterized by $\boldsymbol{\Psi}=\left\{ \mathcal{C}_{n},\bm{\psi}_{n}\right\} _{n=1}^{N}.$
The difference between time-invariant and time-varying channel statistics
model is shown in Fig.~\ref{fig:Channel-statistics-model}. The channel
statistics are assumed to be known to SP. This can be accomplished,
e.g., by building a database of the propagation environment combined
with users sending their planned trajectory to the scheduler.  This
is further elaborated in Section \ref{sec:Channel-Predictability}. 

\subsection{Cost function}

The load is a function of the amount of service $\mathsf{S}_{n}$
that is provided to a user $n$, 
\begin{align}
L_{n} & =\mathsf{S}_{n}.\label{eq:genericLoad}
\end{align}
We can view $\mathsf{S}_{n}$, for example, as the total number of
bits to be delivered. The cost to the SP for serving $N$ users, with
vector of channels $\mathbf{g}=[g_{1},\ldots,g_{N}]^{T}$ is denoted
by $C_{d}(\mathbf{L};\mathbf{g})$, where $C_{d}:(\mathbb{R}_{+}^{N},\mathbb{R}_{+}^{N})\rightarrow\mathbb{R}_{+}$,
is strictly convex and increasing in $\mathbf{L}=[L_{1},\ldots,L_{N}]^{T}$,
while being decreasing in $\mathbf{g}$ (since a better channel require
less cost for a given load $\mathbf{L}$).

We study two network models, reactive and proactive, whose cost functions
are further described below.

\subsubsection{Reactive network model}

The reactive network model will be our baseline approach. In the reactive
network model, the SP has to serve the user data requests upon their
arrival. The amount of load user $n$ generates in time slot $t$
for a reactive network (\ref{eq:genericLoad}) can be computed as

\begin{align}
L_{n,t}^{\mathcal{R}} & =S\,d_{n,t}.
\end{align}
We can write for the reactive network, the time-averaged cost as 
\begin{equation}
c^{\mathcal{R}}(\bar{\boldsymbol{\pi}},\boldsymbol{\Psi}^{Q})=\underset{t\rightarrow\infty}{\limsup}\frac{1}{t}\sum_{l=0}^{t-1}\,\mathbb{E}\biggl[C_{d}\Bigl(\mathbf{L}_{t}^{\mathcal{R}};\mathbf{g}_{t}\Bigr)\biggr],\label{eq:reactive_cost}
\end{equation}
where expectation is over the joint demand and joint channel statistics
of the users, in which as before $\mathbf{L}_{t}^{\mathcal{R}}$ is
the vector of loads and $\mathbf{g}_{t}=[g_{1,t},g_{2,t},\ldots,g_{N,t}]^{T},\,g_{n,t}\in\mathcal{C}_{n}$
is the vector of channels at time $t$. 

\subsubsection{Proactive network model}

\begin{figure}
\centering

\includegraphics[width=1\columnwidth]{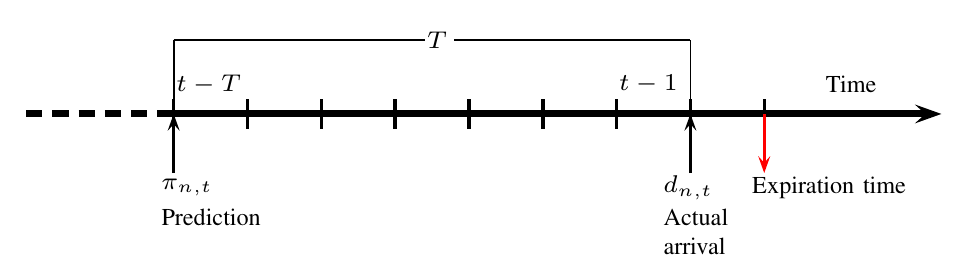}

\caption{\label{fig:Demand-statistics-model}A user request $d_{n,t}$ is proactively
served $T$ (proactive service window) slots ahead, where $\pi_{n,t}$
is the probability of request $d_{n,t}$ being realized at time $t$.}
\end{figure}

Unlike reactive network, the proactive network possesses the flexibility
in servicing the user data requests before their actual realization.
Therefore, the SP utilizes the demand profile $\bar{\boldsymbol{\pi}}$
of the users in providing proactive serve for each request in $T$
time slots ahead, where $T$ denotes the proactive service window
(see Fig. \ref{fig:Demand-statistics-model}). The key factor that
determines the choice of $T$ is content recency and availability,
as users may consume content that is not older than $T$ slots. The
SP knows the demand $\bar{\boldsymbol{\pi}}$ and channel $\boldsymbol{\Psi}^{Q}$
profile of the users, and therefore it tries to even out the load
over the $T$ time-slot proactive service window. To this end, we
denote by $u_{n,t}(\tau)$ the amount of \emph{proactive service}
applied to a user $n$ at time slot $t$ for a possible request, $\tau$
slots in the future, i.e., at time $t+\tau$, where $1\leq\tau\leq T$.\footnote{The notation of the proactive service $u_{n,t}(\tau)$ can best understood
with an example. Consider the case with $t=1$ and $\tau=2$, then
$u_{n,1}(2)$ indicates the proactive service applied in time slot
$1$ for a future possible request in time slot 3, i.e., two slots
ahead of the current time slot.} The proactive service at times $t-\tau$ for a future request at
time $t$ cannot exceed the total demand of $S$ units of service,
i.e.,
\begin{equation}
\sum_{\tau=1}^{T}u_{n,t-\tau}(\tau)\leq S\label{eq:proac_cont_const_1}
\end{equation}
and the proactive service can never be negative, i.e.,
\begin{equation}
u_{n,t}(\tau)\geq0.\label{eq:proac_cont_const_2}
\end{equation}
With the total proactive service at time $t$ denoted by $\mathbf{u}_{n,t}=[u_{n,t}(1),\ldots,u_{n,t}(T)]$,
the load of user $n$ is written as 
\begin{align}
 & L_{n,t}^{\mathcal{\mathcal{P}}}(\mathbf{u}_{n,t})=\label{eq:loadproactive}\\
 & (S-\sum_{\tau=1}^{T}u_{n,t-\tau}(\tau))d_{n,t}+\sum_{\tau=1}^{T}u_{n,t}(\tau).\nonumber 
\end{align}
In \ref{eq:loadproactive}, the term $\sum_{\tau=1}^{T}u_{n,t-\tau}(\tau)$
denotes the amount of proactive services for user $n$ that was already
applied earlier. Therefor in time slot $t$, the remaining data that
has to be served by SP based on the user request is $S-\sum_{\tau=1}^{T}u_{n,t-\tau}(\tau)$.
Finally, the term $\sum_{\tau=1}^{T}u_{n,t}(\tau)$ corresponds to
the proactive service to be applied at time $t$ for user $n$ over
the \emph{next} $T$ slots, to be used in the future. The goal of
the proactive controller is to determine the optimal online proactive
service policy that minimizes the time averaged expected cost while
delivering the content on time: 
\begin{figure*}
\begin{centering}
\subfloat[]{\begin{centering}
\includegraphics[width=0.45\textwidth]{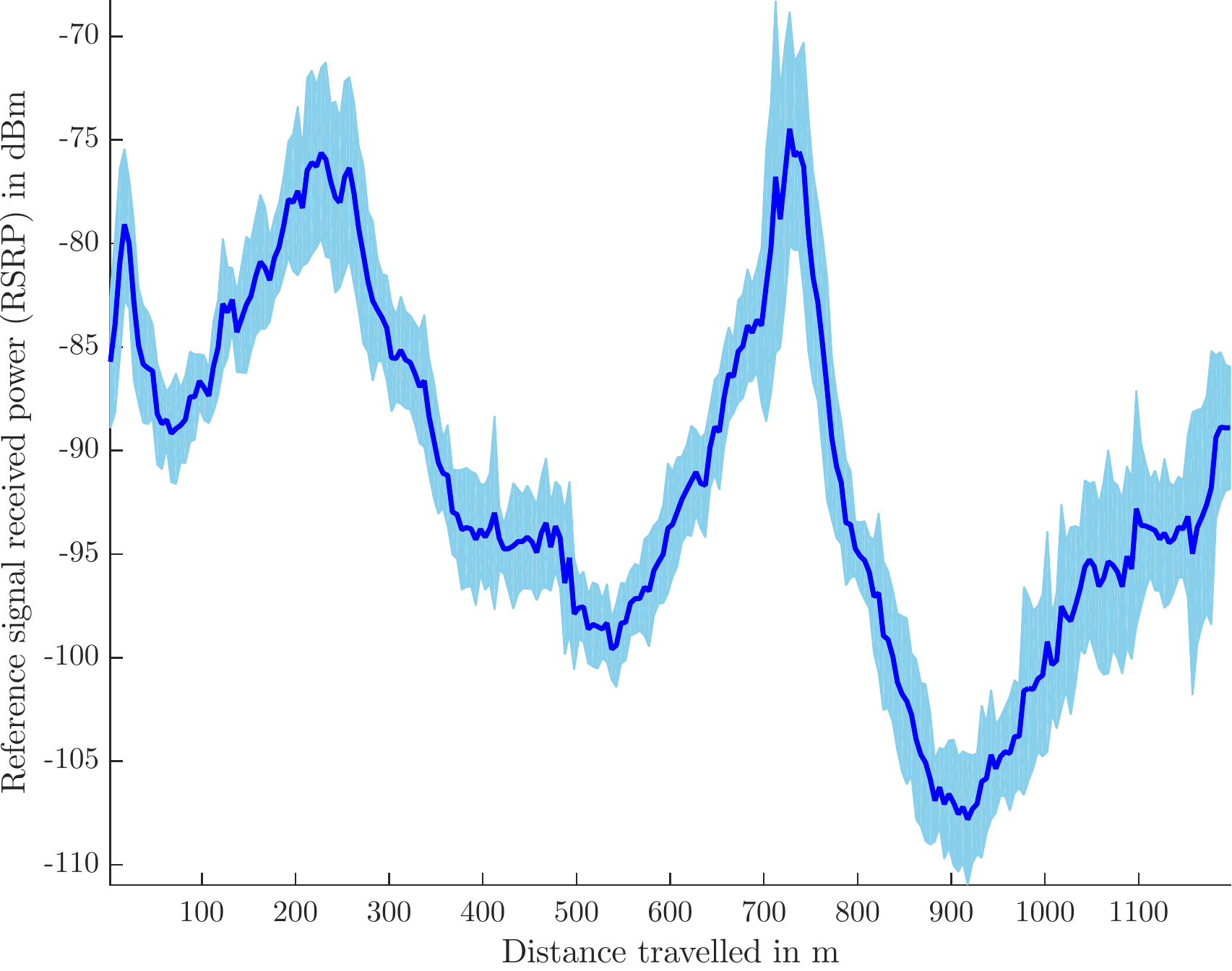} 
\par\end{centering}
}\quad\subfloat[]{\includegraphics[width=0.45\textwidth]{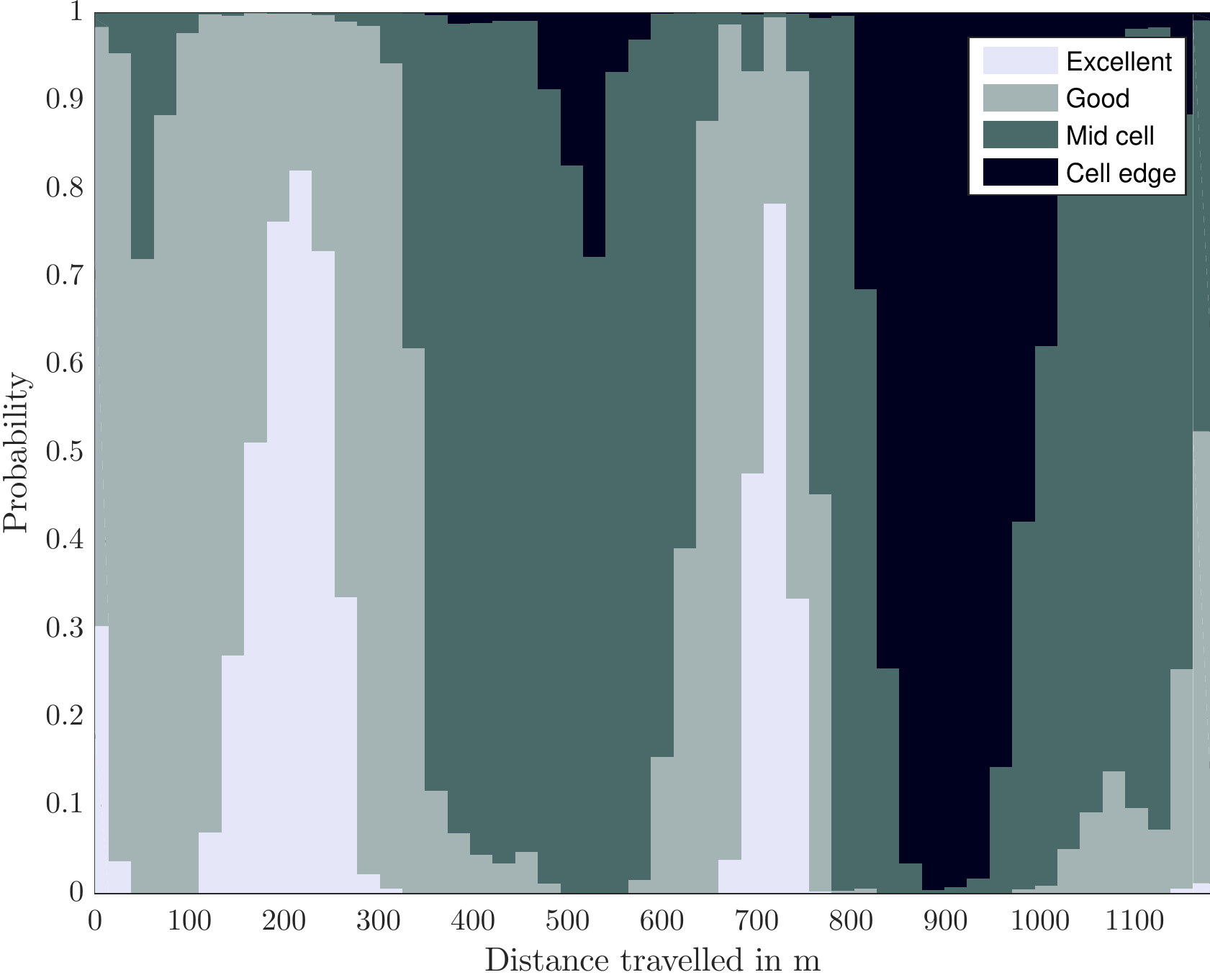}}
\par\end{centering}
\caption{\label{fig:RSRP_two_scenarios}We have conducted channel measurement
campaign using a smart phone, wherein we measured RSRP. Inset (a)
a user is walking along a pedestrian path (Pilbågsgatan-Läraregatan).
The solid line is the mean RSRP with measurements averaged over different
times of the day and at different days. The shaded region captures
the standard deviation of the measurements. Inset (b) depicts the
channel state probabilities along the path. }
\end{figure*}
\begin{eqnarray}
 & \underset{\{u_{n,t}(\tau)\}_{n,t,\tau}}{\min} & \underset{t\rightarrow\infty}{\limsup}\frac{1}{t}\sum_{t'=0}^{t-1}\,\mathbb{E}\biggl[C_{d}\Bigl(\mathbf{L}_{t'}^{\mathcal{P}}(\mathbf{u}_{t'});\mathbf{g}_{t'}\Bigr)\biggr]\nonumber \\
 & \textrm{s.t.} & (\ref{eq:proac_cont_const_1}),\,(\ref{eq:proac_cont_const_2}),\label{eq:proc_optimization_1}
\end{eqnarray}
the optimal value of which is denoted by $c_{T}^{\mathcal{P}}(\bar{\boldsymbol{\pi}},\boldsymbol{\Psi}^{Q})$.
Here we introduced $\mathbf{L}_{t'}^{\mathcal{P}}$ as the vector
of loads and $\mathbf{u}_{t'}=[\mathbf{u}_{1,t'}^{T},\ldots,\mathbf{u}_{N,t'}^{T}]^{T}$.
First, we will derive a global lower bound on the proactive scheduling
cost (\ref{eq:proc_optimization_1}) and then design an asymptotically
optimal policy that achieves the lower bound. Note that the reactive
model is recovered when $u_{n,t}(\tau)\equiv0$. 

\section{Channel Predictability\label{sec:Channel-Predictability}}

The aforementioned proactive model relies on the assumption that channel
statistics can be predicted for an extended period of time. To validate
this assumption, we have performed a measurement campaign using off-the-shelf
hardware in Gothenburg, Sweden \cite{phdthesisSrikar}. Similar findings
have been reported in \cite{kim2011cooperative,malmirchegini2012spatial,mekki2016channel,liao2015channel,mostofi2010estimation}. 

\subsection{Measurement set-up}

The channel quality measurements were in the form of reference signal
received power (RSRP), an important metric for measuring cell selection
and handover. RSRP was collected by using a Google Nexus 5X smartphone
and logged using GNet Track Pro by Gyokov solutions. The application
allowed logging of signal strength, GPS position and many other parameters.
Measurements along pre-specified paths were taken during various times
of the day. These measurements were then mapped to a one dimensional
space with the distance from a fixed point as one of the parameters.
The starting and ending points are located at known fixed geographical
positions, allowing the same track to be recorded several times. Three
measurement campaigns were conducted: one walking on a street, one
in a tram going through a tunnel, and one on a bus touring the city. 

\subsection{Measurement findings}

Fig.~\ref{fig:RSRP_two_scenarios} (a) shows the RSRP values along
the track for the walking path. We observe large variations of the
RSRP, depending on the position. We also see that over the span of
multiple days, the RSRP at a given position is relatively stable,
with variations due to environmental factors as well as GPS errors.
Peaks in the RSRP are due to direct line of sight connections with
base stations, while valleys are due to shadowing by large buildings
and other structures. In conclusion, the assumption of predictable
received power appears to be reasonable, provided the path of the
user along the proactive service window is known. The distributions
from Fig.~\ref{fig:Channel-statistics-model} can be related to Fig.~\ref{fig:RSRP_two_scenarios}
(b), where space is mapped into time slots and the RSRP values are
quantized. The channel measurements in Fig.~\ref{fig:RSRP_two_scenarios}
(a) are categorized\footnote{See http://laroccasolutions.com/164-rsrq-to-sinr/.}
in four channel states namely, excellent ($\textrm{RSRP}\geq-80\,\textrm{dBm}$)
, good ($-80\,\textrm{dBm}<\textrm{RSRP}\leq-90\,\textrm{dBm}$),
mid cell ($-90\,\textrm{dBm}<\textrm{RSRP}\leq-100\,\textrm{dBm}$),
and cell edge ($\textrm{RSRP}\leq-100\text{\,\textrm{dBm}}$). In
Fig.~\ref{fig:RSRP_two_scenarios} (b), we can clearly observe channel
state probabilities varying with position. Depending on the velocity
of the user, the channel state probabilities can be mapped to time.
It is clear that the measurements follow the time-varying model in
Fig.~\ref{fig:Channel-statistics-model} (b), rather than the time-invariant
model from Fig.~\ref{fig:Channel-statistics-model} (a). We note
that the corresponding average time-invariant probabilities are {[}0.11
0.32 0.41 0.16{]}, from excellent to cell edge channel conditions.

\section{Proactive Service with Future Channel Statistics\label{sec:Proactive-TITI}}

In this section, we aim to find solutions to (\ref{eq:proc_optimization_1}),
leading to proactive gains to flatten the load and cost. As the infinite
horizon optimization of (\ref{eq:proc_optimization_1}) is of infinite
dimensionality, its exact solution is computationally intractable.
We thus propose to tackle this problem through a two-step approach.
First, we establish a fundamental lower bound on the minimum cost,
then we develop an asymptotically optimal proactive caching strategy
that achieves the lower bound as the proactive service window grows.

Given $\mathbf{g}_{t}$, a channel realization vector in time $t$,
the probability of this vector being realized is $P_{c}(\mathbf{g}_{t})$. 

\subsection{Time-invariant channel statistics\protect\footnote{This section is largely a summary of \cite{muppirisetty2015proactive},
but will help the reader to understand the more complex case with
time-varying channel statistics.}}

We establish a global lower bound on the achievable costs by any proactive
caching policy. We first introduce the following notation: $\mathcal{B}_{t}=\{n\in\mathcal{N}:\;d_{n,t}=1\}$
with probability $P_{d}(\mathcal{B})$, the set of requesting users
at time $t$, and $P_{c}(\mathbf{g})$ the probability of an aggregate
channel $\mathbf{g}=[g_{1},\ldots,g_{N}]$. As the channel statistics
are time-invariant, therefore $P_{c}(\mathbf{g})$ does not depend
on $t$ and can be easily derived from $\boldsymbol{\Psi}$. 
\begin{thm}
\label{thm:TITI}The optimal time average expected cost $c_{T}^{\mathcal{P}}(\bar{\boldsymbol{\pi}},\boldsymbol{\Psi})$
satisfies 
\begin{equation}
c_{T}^{\mathcal{P}}(\bar{\boldsymbol{\pi}},\boldsymbol{\Psi})\geq\underline{c}_{\mathcal{U}}(\bar{\boldsymbol{\pi}},\boldsymbol{\Psi}),
\end{equation}
where $\underline{c}_{\mathcal{U}}(\bar{\boldsymbol{\pi}},\boldsymbol{\Psi})$
is the optimal value of 
\begin{align}
 & \min\Biggl\{\sum_{\mathbf{g}\in\mathcal{C}}\sum_{\mathcal{B}\subseteq\mathcal{N}}P_{c}(\mathbf{g})P_{d}(\mathcal{B})\nonumber \\
 & C_{d}\Biggl(\left[(S-\bar{\mu}_{n})+\tilde{\mu}_{n}\left(\mathcal{B},\mathbf{g}\right)\right]_{n=1}^{N};\mathbf{g}\Biggr)\Biggr\}\nonumber \\
 & \text{\textrm{}subject to \ensuremath{0\leq\tilde{\mu}_{n}\left(\mathcal{B},\mathcal{\mathbf{g}}\right)\leq S},}\label{eq:proactive_cost}
\end{align}
where $\tilde{\mu}_{n}\left(\mathcal{B},\mathcal{\mathbf{g}}\right)$
are the optimization variables and $\bar{\mu}_{n}=\sum_{\mathbf{h}\in\mathcal{C}}\sum_{\mathcal{D}\subseteq\mathcal{N}}P_{c}(\mathbf{h})P_{d}(\mathcal{D})\,\tilde{\mu}_{n}\left(\mathcal{D},\mathcal{\mathbf{h}}\right)$. 
\end{thm}
\begin{IEEEproof}
See Appendix \ref{sec:Lower-bound-proof}. 
\end{IEEEproof}

\subsubsection*{Remark}

In the objective of (\ref{eq:proactive_cost}), the term $\bar{\mu}_{n}$
represents the average of the proactively cached content at user $n$
for an expected request. The term $\tilde{\mu}_{n}\left(\mathcal{B},\mathbf{g}\right)$
is the average of the total amount of proactively cached content at
user $n$ in a time slot, when the set of requesting users is $\mathcal{B}$
and the respective channel gain is $\mathbf{g}$. Now, observing that
the optimization problem of $\underline{c}_{\mathcal{U}}(\bar{\boldsymbol{\pi}},\boldsymbol{\Psi})$
is convex with closed and bounded constraints set, the optimization
problem has a unique solution. 

We now harness the established lower bound and the optimization (\ref{eq:proactive_cost}),
to develop our proposed asymptotically optimal stationary policy
\begin{defn}
(Policy $\wp_{\mathcal{U}}$) Let $\{\tilde{\mu}_{n}\left(\mathcal{B},\mathcal{\mathbf{g}}\right)\}_{n,\mathcal{B},\mathbf{g}}$
be the optimal solution to \eqref{eq:proactive_cost}. We consider
the proactive scheduling policy $\wp_{\mathcal{U}}$ that in every
time slot observes the set of requesting users $\mathcal{B}_{t}$,
and channel gain realization $\mathbf{g}_{t}$, and then decides a
proactive caching control $u_{n,t}(\tau)=\frac{1}{T}\tilde{\mu}_{n}\left(\mathcal{B}_{t},\mathcal{\mathbf{g}}_{t}\right)$,
$\forall n,t,1\leq\tau\leq T$. 
\end{defn}
The policy $\wp_{\mathcal{U}}$ is determined offline, based on the
demand $\bar{\boldsymbol{\pi}}$ and channel $\boldsymbol{\Psi}$
profiles. During online operation, the policy is a function of the
current realization of demand and channel and requires a look-up table
of length $2^{N}\prod_{n}|\mathcal{C}_{n}|$, which entails a search
process of complexity $O(N+\sum_{n}\log(|\mathcal{C}_{n}|))$. Note
that, to apply policy $\wp_{\mathcal{U}}$, the solution of \eqref{eq:proactive_cost}
has to be obtained offline based on the demand $\bar{\boldsymbol{\pi}}$
and channel $\boldsymbol{\Psi}$ profiles. We can now establish the
asymptotic optimality property of policy $\wp_{\mathcal{U}}$.
\begin{thm}
\label{th:asymptotic} Denote the time average expected cost under
policy $\wp_{\mathcal{U}}$ by $c_{T}^{\wp_{\mathcal{U}}}(\bar{\boldsymbol{\pi}},\boldsymbol{\Psi})$.
Then policy $\wp_{\mathcal{U}}$ is asymptotically optimal, in the
sense that 
\[
\limsup_{T\to\infty}|c_{T}^{\wp_{\mathcal{U}}}(\bar{\boldsymbol{\pi}},\boldsymbol{\Psi})-c_{T}^{\mathcal{P}}(\bar{\boldsymbol{\pi}},\boldsymbol{\Psi})|=0.
\]
\end{thm}
\begin{IEEEproof}
See Appendix \ref{app:asymptotic}. 
\end{IEEEproof}
With the relevant characteristics of proactive caching for time-invariant
channels have been investigated, we are ready to consider the scenario
of of time-varying channel statistics.

\subsection{Time-varying channel statistics\label{sec:Proactive-TITV}}

In this section, we follow a similar procedure, but for the case with
time-varying channel statistics. Due to the cyclo-stationary nature
of the channel, we introduce a new random variable, in order to develop
a stationary policy. We denote by $s_{t}\in\mathcal{Q}=\{0,1,\ldots,Q-1\}$
the index in the period corresponding to time slot $t$, with $P_{\mathrm{s}}(s_{t})=1/Q$.
The channel statistics can thus be interpreted as a function of $s$.
\begin{thm}
\label{thm:TITV}When $T$ is an integer multiple\footnote{The general case of arbitrary $T$ is treated in Appendix \ref{sec:Proof-of-Theorem_TITV}
and leads to somewhat less elegant expressions.} of $Q$, the minimum time average cost $c_{T}^{\mathcal{P}}(\bar{\boldsymbol{\pi}},\boldsymbol{\Psi})$
satisfies  
\begin{equation}
c_{T}^{\mathcal{P}}(\bar{\boldsymbol{\pi}},\boldsymbol{\Psi}^{Q})\geq\underline{c}_{\mathcal{F}}(\bar{\boldsymbol{\pi}},\boldsymbol{\Psi}^{Q}),
\end{equation}
where $\underline{c}_{\mathcal{F}}(\bar{\boldsymbol{\pi}},\boldsymbol{\Psi}^{Q})$
is the optimal value of 
\begin{align}
 & \min\Biggl\{\sum_{s\in\mathcal{Q}}P_{\mathrm{s}}(s)\sum_{\mathbf{g}\in\mathcal{C}}\,P_{c}(\mathbf{g}\vert s)\sum_{\mathcal{B}\subseteq\mathcal{N}}\,P_{d}(\mathcal{B})\nonumber \\
 & \times C_{d}\biggl(\Bigl[\delta_{n\in\mathcal{B}}(S-\bar{\mu}_{n}(s))\nonumber \\
 & +\sum_{s'\in\mathcal{Q}}P_{\mathrm{s}}(s')\tilde{\mu}_{n}\left(\mathcal{B},\mathbf{g},s,s'\right)\Bigr]_{n=1}^{N};\mathbf{g}\biggr)\Biggr\}\nonumber \\
 & \text{\textrm{}subject to \ensuremath{0\leq\tilde{\mu}_{n}\left(\mathcal{B},\mathbf{g},s,s'\right)\leq S}},\label{eq:proactive_cost_TITV}
\end{align}
where $\tilde{\mu}_{n}\left(\mathcal{B},\mathbf{g},s,s'\right)$ are
the optimization variables and
\begin{align*}
 & \bar{\mu}_{n}(s)=\\
 & \sum_{s'\in\mathcal{Q}}P_{\mathrm{s}}(s')\,\sum_{\mathbf{h}\in\mathcal{C}}P_{c}(\mathbf{h}|s')\sum_{\mathcal{D}\subseteq\mathcal{N}}P_{d}(\mathcal{D})\,\tilde{\mu}_{n}\left(\mathcal{D},\mathcal{\mathbf{h}},s',s\right).
\end{align*}
\end{thm}
\begin{IEEEproof}
Please refer Appendix \ref{sec:Proof-of-Theorem_TITV}.
\end{IEEEproof}

\subsubsection*{Remark}

Similar to the optimization of $\underline{c}_{\mathcal{U}}(\bar{\boldsymbol{\pi}},\boldsymbol{\Psi})$,
the optimization $\underline{c}_{\mathcal{F}}(\bar{\boldsymbol{\pi}},\boldsymbol{\Psi}^{Q})$
is also convex with a unique solution since the constraints set is
compact and the objective function $C_{d}(.)$ is strictly convex.
In the objective of (\ref{eq:proactive_cost_TITV}) the term $\bar{\mu}_{n}(s)$
captures the average proactive service offered to the user $n$ before
the actual demand request has arrived, when the current time slot
corresponds to index $s$. Further, the term $\sum_{s'\in\mathcal{Q}}P_{\mathrm{s}}(s')\tilde{\mu}_{n}\left(\mathcal{B},\mathbf{g},s,s'\right)$
captures the expected amount of content proactively served to user
$n$, when the set of demanding users is $\mathcal{B}$, the current
slot corresponds to index $s$, and the present channel realization
is $\mathbf{g}$. 

We now show an asymptotically optimal policy design that attains the
lower bound (\ref{eq:proactive_cost_TITV}). 
\begin{defn}
(Policy $\wp_{\mathcal{F}}$) Given the current set $\mathcal{B}_{t}$
of requesting users in a time slot $t$, channel gain realization
$\mathbf{g}_{t}$, and the current index in the period $s_{t}$, then
the proactive scheduler $\wp_{\mathcal{F}}$ assigns proactive controls
as $u_{n,t}(\tau)=\frac{1}{T}\tilde{\mu}_{n}\left(\mathcal{B}_{t},\mathbf{g}_{t},s_{t},s_{t+\tau}\right),$
$\forall n,t,1\leq\tau\leq T$, where $\tilde{\mu}_{n}\left(\mathcal{B},\mathbf{g},s,s'\right)$
denote the optimal solution to (\ref{eq:proactive_cost_TITV}).

The proposed policy $\wp_{\mathcal{F}}$ requires a look-up table
of length $2^{N}\prod_{n}|\mathcal{C}_{n}|\,|\mathcal{Q}|^{2}$, that
matches each combination of a requesting set of users and channel
gain realization with a proactive caching control. The complexity
of searching such a look-up table is $O(N+\sum_{n}\log(|\mathcal{C}_{n}|)+2\,\log(|\mathcal{Q}|))$.
In comparison to $\wp_{\mathcal{U}}$, the policy $\wp_{\mathcal{F}}$
takes in to consideration statistical information not only of the
current channel probabilities (represented by $s_{t}$) but also the
future set of channel probabilities through $s_{t+\tau},\tau=1,\ldots,T$.
This will help scheduler to shift the loads to time slots where the
probabilities of higher channel state values are higher in order to
minimize the overall network cost. 
\end{defn}
\begin{thm}
Under the time-invariant demand and time-varying channel statistics
model described by $\boldsymbol{\Psi}^{Q}$. The policy $\wp_{\mathcal{F}}$
is asymptotically optimal, in the sense that $\limsup_{T\to\infty}$
$|c_{T}^{\wp_{\mathcal{F}}}(\bar{\boldsymbol{\pi}},\boldsymbol{\Psi}^{Q})-$
$c_{T}^{\mathcal{P}}(\bar{\boldsymbol{\pi}},\boldsymbol{\Psi}^{Q})|$
$=0$, where $c_{T}^{\wp_{\mathcal{F}}}(\bar{\boldsymbol{\pi}},\boldsymbol{\Psi}^{Q})$
denotes the time average expected cost under policy $\wp_{\mathcal{F}}$.
\end{thm}
\begin{IEEEproof}
The proof is similar to that of Theorem \ref{th:asymptotic}, and
omitted here. 
\end{IEEEproof}

\section{Numerical Results and Discussion\label{sec:Numerical-Results}}

We assume that the network scheduler is aware of the user demand $\bar{\boldsymbol{\pi}}$
and channel $\boldsymbol{\Psi}^{Q}$ profiles. The scheduler spends
$S=1$ units of service for each request. We take the cost function
for the demand to be of a simple polynomial form $C_{d}(\mathbf{L}_{t},\mathbf{g}_{t})=\sum_{n=1}^{N}L_{n,t}^{4}/g_{n,t}$.
While this choice of cost function is arbitrary, it is meant merely
to illustrate the behavior of the proactive scheduler. Since demands
will always be satisfied, performance will be evaluated in terms of
the expected cost and the expected load. 

\subsection{Time-invariant demand and channel statistics }

We assume each user $n$ observes one of the two possible channel
states $\{g_{n}^{(1)},g_{n}^{(2)}\}$ with probabilities $\{\psi_{n}^{(1)},\psi_{n}^{(2)}=1-\psi_{n}^{(1)}\}$.
We consider $g_{n}^{(2)}\geq g_{n}^{(1)}$, hence $g_{n}^{(2)}$ is
termed the good channel state, while $g_{n}^{(1)}$ is the bad channel
state. Furthermore, for convenience we assume same channel state values
and corresponding channel probability values for all the users.

\subsubsection{Impact of demand and channel probabilities on the expected cost}

\begin{figure}
\includegraphics[width=1\columnwidth]{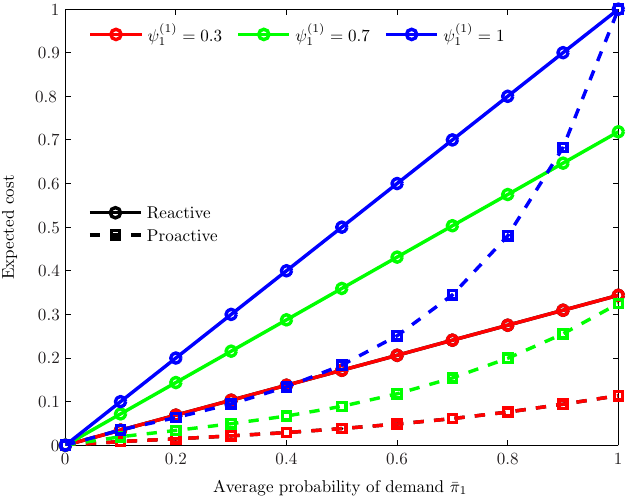}\caption{\label{fig:expected_cost}The expected cost for reactive and proactive
schemes for single user scenario under the time-invariant demand and
channel statistics model. The probabilities of the demand $\bar{\pi}_{1}$
and channel $\psi_{1}^{(1)}$ are varied with the channel states fixed
to $g_{1}^{(1)}=1$ (bad channel state), $g_{1}^{(2)}=2$ (good channel
state).}
\end{figure}

In Fig. \ref{fig:expected_cost}, we compare the expected cost achieved
by proactive and reactive schemes under time-invariant demand and
channel statistics model for a single user scenario. We can infer
easily from the Fig. \ref{fig:expected_cost} that the proactive schemes
always offer lower expected cost compared to the reactive scheme irrespective
of any demand and channel probability setting $(\bar{\pi}_{1},\psi_{1}^{(1)})$.
The main reason for higher expected cost for reactive scheme is that,
the reactive scheme does not possess the flexibility to delay the
service requests and it has to fulfill the demand requests whenever
they are initiated from the user. On the other hand the proactive
scheme offers flexibility in the scheduling strategy by exploiting
the demand and channel statistics to minimize cost by load balancing.
When the demand $\bar{\pi}_{1}$ and channel $\psi_{1}^{(1)}$ probabilities
are increased, the expected cost offered by both reactive and proactive
is also increased. The cost increases with increase in $\bar{\pi}_{1}$
due to the fact that the system is more loaded with incoming demand
requests. The reason for the cost to increase with $\psi_{1}^{(1)}$
is as the user more often experiences a bad channel state $g_{1}^{(1)}$
than a good channel state $g_{1}^{(2)}$.

In \cite{John2014}, it was shown that both reactive and proactive
schemes converge when $\bar{\pi}_{1}=1,$ under time-invariant demand
statistics scenario and no channel knowledge. However, from Fig. \ref{fig:expected_cost},
we can observe that both schemes does not converge for $\bar{\pi}_{1}=1$.
We can show easily that expected cost for both schemes is same when
the user always observes either the good or bad channel state all
the time. So, it can be concluded that there is advantage in apply
proactive service, if there is no variation of channels and the demand
is certain. However, when the channels vary from one slot to another
(i.e., $0<\psi_{1}^{(1)}<1$), then even with certain data demand
there is still potential to apply proactive service in the presence
of good channel so as to minimize the cost when the bad channel is
realized.

\subsubsection{Impact of the value of channel states on the expected cost}

\begin{figure}
\includegraphics[width=1\columnwidth]{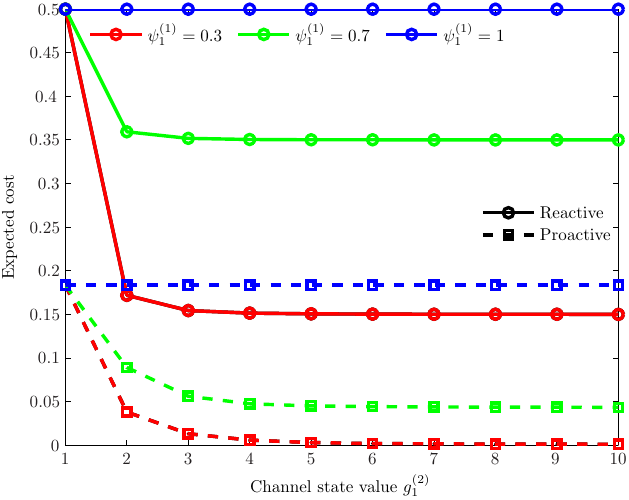}\caption{\label{fig:impact_on_g2}The expected cost for reactive and proactive
costs for a single user scenario for time-invariant demand and channel
statistics model. The channel probability $\psi_{1}^{(1)}$ and channel
state $g_{1}^{(2)}$ are varied with fixed user demand probability
to $\bar{\pi}_{1}=0.5$ and fixed $g_{1}^{(1)}=1$.}
\end{figure}

The impact of channel state on the expected cost for a single user
scenario is depicted in Fig. \ref{fig:impact_on_g2}. For this scenario,
the good channel state $g_{1}^{(2)}$ is increased while the other
bad channel state $g_{1}^{(1)}=1$ is kept constant, and the demand
probability is set to $\bar{\pi}_{1}=0.5$.  When $\psi_{1}^{(1)}=1$
(shown in blue), which means the user always observes $g_{1}^{(1)}$,
there is no impact of $g_{1}^{(2)}$ on the expected cost for both
the schemes. For $\psi_{1}^{(1)}=0.3$ and $\psi_{1}^{(1)}=0.7$,
the cost decreases with increase in $g_{1}^{(2)}$. This is expected,
as when one of the channel states becomes good, the applied proactive
service is shifted to that channel condition to minimize the cost.

One interesting observation is the decrease in expected cost is significant
when when $g_{1}^{(2)}$ is twice $g_{1}^{(1)}$, while beyond this
point the reduction in cost is minimal. This phenomenon is related
to choice of the fourth-order polynomial for the cost function. It
is expected that expected cost will slowly decrease when $g_{1}^{(2)}>2$
for lower-order polynomial cost functions. 

\subsubsection{Impact of proactive service window size on the expected cost}

\begin{figure}
\centering

\includegraphics[width=1\columnwidth]{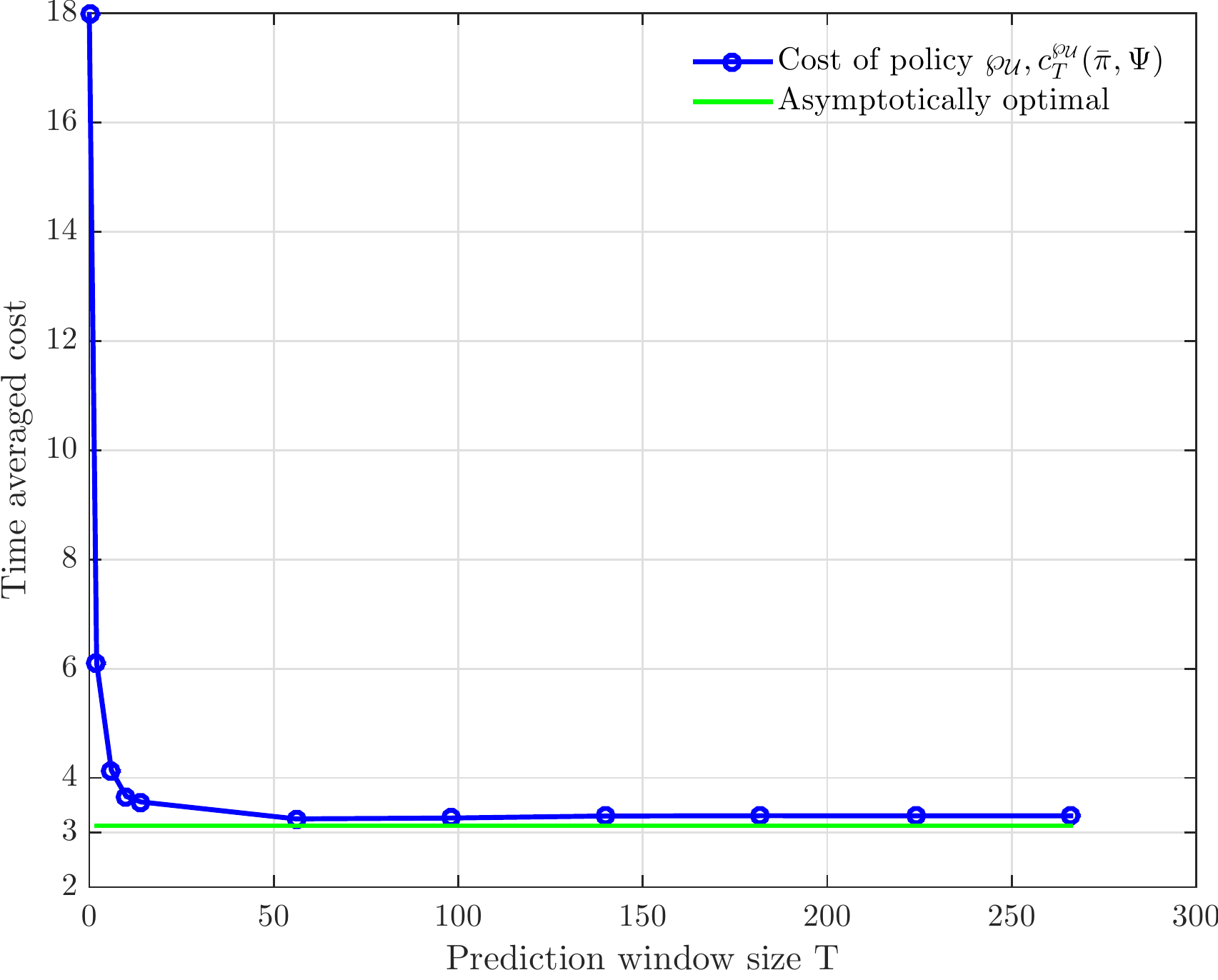}\caption{\label{fig:impact_on_T}Impact of proactive window size $T$ on achievable
cost for time-invariant demand and channel statistics model. The results
have been averaged over 40 simulations and for $t=10000$ time slots.}

\end{figure}

We consider a two-user scenario with demand probability set to $\bar{\pi}_{n}=0.42,n=1,2$,
and channel state probabilities set to $\psi_{n}^{(1)}=0.54,n=1,2$,
and with same channel state values $g_{n}^{(1)}=0.5,g_{n}^{(2)}=2$,
for both users. Under this setting the time averaged cost $\wp_{\mathcal{U}}$
is evaluated when the proactive service window size $T$ is increased.
In Fig. \ref{fig:impact_on_T}, we plot the cost of policy $\wp_{\mathcal{U}}$
against asymptotically optimal limit under time-invariant demand and
channel statistics. The policy $\wp_{\mathcal{U}}$ converges quickly
with $T$ (especially after $T=50$) to the established lower bound
$\underline{c}_{\mathcal{U}}(\bar{\boldsymbol{\pi}},\boldsymbol{\Psi})$.
To draw more insights, we move on to the  more realistic case of time-varying
channel statistics in the next subsection.

\subsection{Time-invariant demand and time-varying channel statistics}

We consider $N=2$ users and we set demand probability $\bar{\pi}_{n}=0.42,\forall n$
for this scenario in the system. For the time-varying channel statistics,
we consider $Q=14$ different channel state probability levels for
each state. Similar to earlier numerical results, we consider again
two channel states. We set the channel states to $\{g_{n}^{(1)}=0.5,g_{n}^{(2)}=2\}$. 

\subsubsection{Impact of proactive service window size on the expected cost}

\begin{figure}
\centering

\includegraphics[width=1\columnwidth]{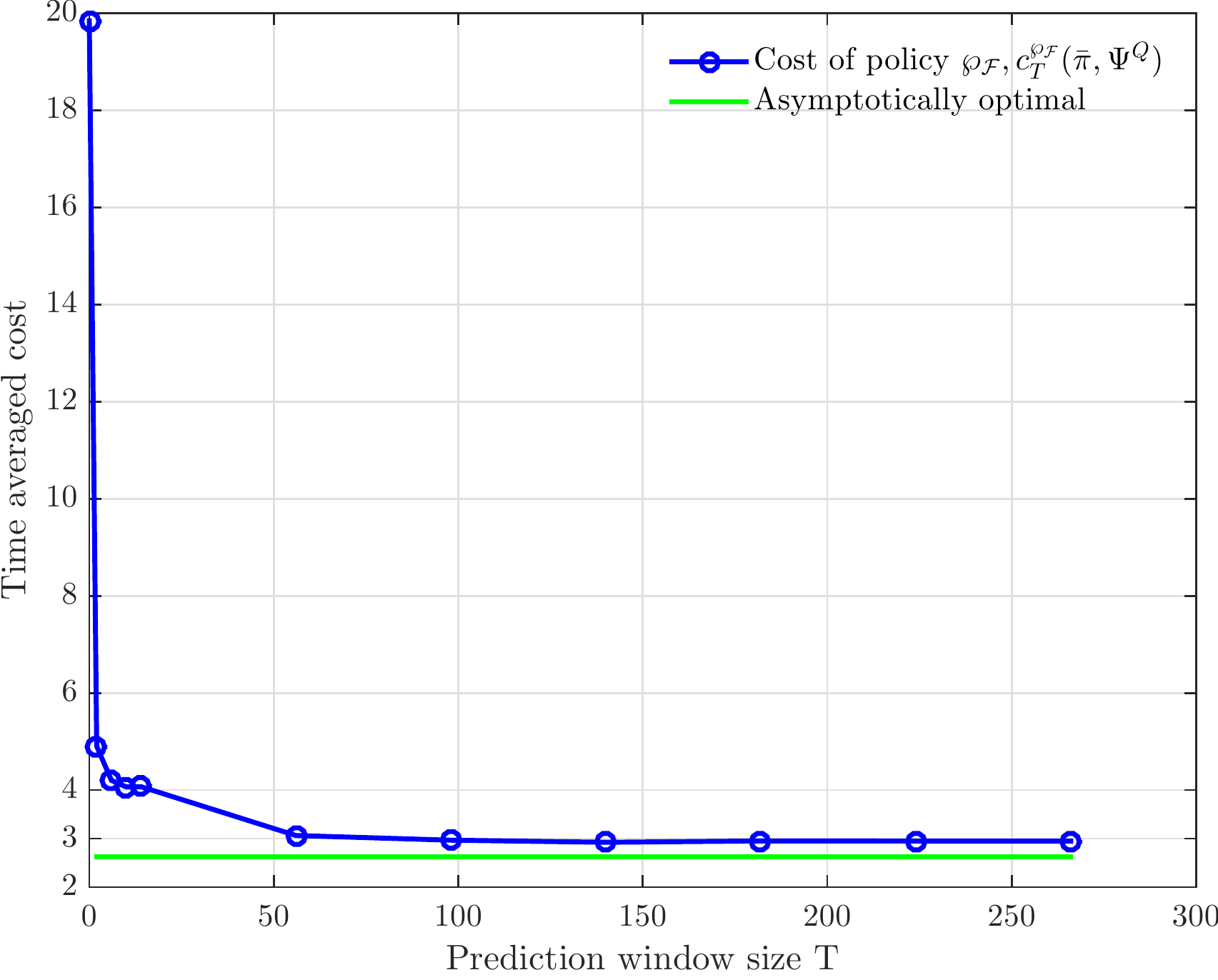}\caption{\label{fig:Impact_on_TITV}Impact of proactive window size $T$ on
achievable cost for time-invariant demand and time-varying channel
statistics model. The results have been averaged over 40 simulations
and for $t=10000$ time slots.}
\end{figure}

The time-varying channel probabilities for the channel state $g_{n}^{(1)}$
are (0.8, 0.9, 0.12, 0.24, 0.89, 0.64, 0.9, 0.11, 0.2, 0.27, 0.89,
0.70, 0.59, 0.14). We further assume the channel profile is same for
all the users. In Fig. \ref{fig:Impact_on_TITV}, we depict the cost
convergence of the policy $\wp_{\mathcal{F}}$ with respect to $\underline{c}_{\mathcal{F}}(\bar{\boldsymbol{\pi}},\boldsymbol{\Psi}^{Q})$
with increase in proactive service window size $T$ under time-invariant
demand and time-varying channel statistics. We can clearly observe
from the plot that the designed policy $\wp_{\mathcal{F}}$ converges
very quickly to the global lower bound $\underline{c}_{\mathcal{F}}(\bar{\boldsymbol{\pi}},\boldsymbol{\Psi}^{Q})$
within proactive service window size of $T=80$. It should be noted
that the value of $\underline{c}_{\mathcal{F}}(\bar{\boldsymbol{\pi}},\boldsymbol{\Psi}^{Q})$
is lower compared to $\underline{c}_{\mathcal{U}}(\bar{\boldsymbol{\pi}},\boldsymbol{\Psi})$
(see Fig. \ref{fig:impact_on_T}) for similar settings due to more
certain channel statistics. 

\subsubsection{Average load and cost levels of different channel probability levels}

\begin{figure*}
\begin{centering}
\subfloat[]{\begin{centering}
\includegraphics[width=0.45\textwidth]{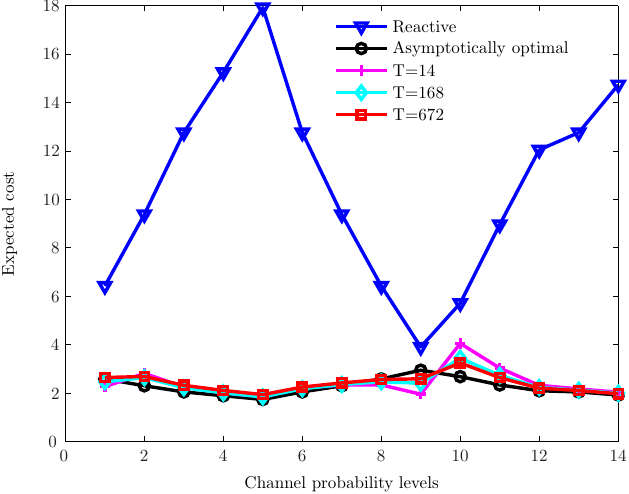} 
\par\end{centering}
}\quad\subfloat[]{\begin{centering}
\includegraphics[width=0.45\textwidth]{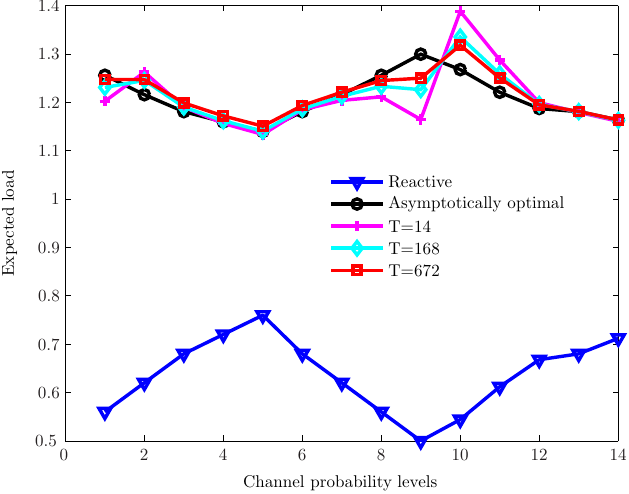} 
\par\end{centering}
}
\par\end{centering}
\caption{\label{fig:avg_load_cost_time_variant_chan_stat}Inset (a) Average
cost levels under reactive and proactive services for time-varying
channel statistics. Inset (b) Average load levels under reactive and
proactive services for time-varying channel statistics.}
\end{figure*}
In Fig. \ref{fig:avg_load_cost_time_variant_chan_stat}, we show the
average cost and load levels for different time periods. For this
case, we set the channel probabilities of the channel state $g_{n}^{(1)}$
as (0.4,0.55,0.7,0.8,0.9,0.7,0.55,0.4,0.25,0.36,0.53,0.67,0.7,0.78).
We compare proactive scheduler $\wp_{\mathcal{F}}$ for various proactive
service window sizes ($T=14,168,672$) against reactive on-time service
and asymptotically optimal limit. We can observe in Fig. \ref{fig:avg_load_cost_time_variant_chan_stat}
(a), that the cost of the reactive service varies considerably with
changing statistics over time. Reactive service does not possess the
flexibility and has to offer service irrespective of the channel conditions.
However, it can be observed that the average cost offered by the proactive
scheduler is constant. This is due to the fact that the proactive
scheduler exhibits flexibility in scheduling and shifts the loads
based on the channel conditions. This can be seen in Fig. \ref{fig:avg_load_cost_time_variant_chan_stat}
(b), where the load for the proactive scheduler is less when the channel
conditions are worse (i.e., channel time periods $q=1,2,3,4,5$).
Furthermore, the load and cost levels of the proactive scheduler approach
to corresponding asymptotically optimal limits with increase in $T$.
For $T=672$, the proactive scheduler load and cost approach to that
of asymptotically optimal (see Fig. \ref{fig:avg_load_cost_time_variant_chan_stat}
(a) and (b)).

\section{Conclusions\label{sec:Conclusions}}

We studied the impact of demand and channel uncertainties on the design
of a proactive scheduler under two scenarios (i) time-invariant demand
and channel statistics, (ii) time-invariant demand and time-varying
channel statistics. Specifically, we established non-trivial global
lower bounds for the two considered scenarios. We showed how to design
cyclo-stationary  asymptotically optimal proactive service policies
that approach such bounds as proactive service window size grow. We
observed that the designed proactive resource scheduler provides better
performance in terms of lower achievable cost, compared to reactive
scheduler. With proactive service, the scheduler has more flexibility
to optimize its loads over time depending on the demand and channel
levels.

\appendices{}

\section{Proof of Theorem \ref{thm:TITI}\label{sec:Lower-bound-proof} }

The optimal value, assuming an optimal policy, is 
\begin{equation}
c_{T}^{\mathcal{P}}(\bar{\boldsymbol{\pi}},\boldsymbol{\Psi})=\underset{t\rightarrow\infty}{\limsup}\frac{1}{t}\sum_{l=0}^{t-1}\,\mathbb{E}\biggl[C_{d}\bigl(\mathbf{L}_{l}^{\mathcal{\mathcal{P}}}(\mathbf{u}_{l},\mathbf{g}_{l})\bigr)\biggr]\label{eq:proc_obj}
\end{equation}
and involves an expectation over all possible sets of requesting users
$\mathcal{B}_{l}$ and their channel state realizations $\mathbf{g}_{l}$
at time $l\geq0$, with associated distribution $P(\mathcal{B}_{l}=\mathcal{B},\mathbf{g}_{l}=\mathbf{g})=P(\mathbf{g}_{l}=\mathbf{g})P(\mathcal{B}_{l}=\mathcal{B})$.
Note that both $P(\mathbf{g}_{l}=\mathbf{g})$ and $P(\mathcal{B}_{l}=\mathcal{B})$
are independent of the time $l$. This allows us to write, after substitution
of (\ref{eq:loadproactive}) into $L_{n,l}^{\mathcal{\mathcal{P}}}(\mathbf{u}_{n,l},g_{n,l})$:
 
\begin{align}
c_{T}^{\mathcal{P}}(\bar{\boldsymbol{\pi}},\boldsymbol{\Psi}) & =\underset{t\rightarrow\infty}{\limsup}\frac{1}{t}\sum_{l=0}^{t-1}\sum_{\mathbf{g}\in\mathcal{C}}\sum_{\mathcal{B}\subseteq\mathcal{N}}P_{c}(\mathbf{g})P_{d}(\mathcal{B})\nonumber \\
 & \times\mathbb{E}\biggl[C_{d}\Bigl(\Bigl[\Bigl(S-\sum_{\tau=1}^{T}u_{n,l-\tau}(\tau)\nonumber \\
 & +\sum_{\tau=1}^{T}u_{n,l}(\tau)\Bigr)\Bigr]_{n=1}^{N};\mathbf{g}\Bigr)\Bigr|\mathcal{B},\mathbf{g}\biggr].\label{eq:jensen}
\end{align}
Since $C_{d}(\cdot)$ is strictly convex, we can use Jensen's inequality
(i.e., $\mathbb{E}[C_{d}(\cdot)]\ge C_{d}(\mathbb{E}[\cdot])$). The
term $\sum_{\tau=1}^{T}u_{n,l-\tau}(\tau)$ is independent on the
current channel $\mathbf{g}$ and set $\mathcal{B}$, yielding 
\begin{align}
c_{T}^{\mathcal{P}}(\bar{\boldsymbol{\pi}},\boldsymbol{\Psi}) & \geq\underset{t\rightarrow\infty}{\limsup}\frac{1}{t}\sum_{l=0}^{t-1}\sum_{\mathbf{g}\in\mathcal{C}}\sum_{\mathcal{B}\subseteq\mathcal{N}}P_{c}(\mathbf{g})P_{d}(\mathcal{B})\nonumber \\
 & \times C_{d}\biggl(\Bigl[\Bigl(S-\sum_{\tau=1}^{T}\mathbb{E}[u_{n,l-\tau}(\tau)]\Bigr)\delta_{n\in\mathcal{B}}\nonumber \\
 & +\sum_{\tau=1}^{T}\mathbb{E}[u_{n,l}(\tau)|\mathcal{B},\mathbf{g}]\Bigr]_{n=1}^{N};\mathbf{g}\biggr)\biggr].
\end{align}
We now push $\underset{t\rightarrow\infty}{\limsup}\frac{1}{t}\sum_{l=0}^{t-1}$
through the summation and through $C_{d}(\cdot)$, using Jensen's
inequality again. Since $\limsup_{t\rightarrow\infty}(-f(t))=-\liminf_{t\rightarrow\infty}(f(t))$,
we have 
\begin{align}
c_{T}^{\mathcal{P}}(\bar{\boldsymbol{\pi}},\boldsymbol{\Psi}) & \geq\sum_{\mathbf{g}\in\mathcal{C}}\sum_{\mathcal{B}\subseteq\mathcal{N}}P_{c}(\mathbf{g})P_{d}(\mathcal{B})\,C_{d}\biggl(\Bigl[\delta_{n\in\mathcal{B}}\Bigl(S\nonumber \\
 & -\underset{t\rightarrow\infty}{\liminf}\frac{1}{t}\sum_{l=0}^{t-1}\sum_{\tau=1}^{T}\mathbb{E}[u_{n,l-\tau}(\tau)]\Bigr)\nonumber \\
 & +\underset{t\rightarrow\infty}{\limsup}\frac{1}{t}\sum_{l=0}^{t-1}\sum_{\tau=1}^{T}\mathbb{E}[u_{n,l}(\tau)|\mathcal{B},\mathbf{g}]\Bigr]_{n=1}^{N};\mathbf{g}\biggr).
\end{align}
As $C_{d}(x)$ is monotonically increasing in $x$, replacing $\limsup$
on the right hand side of the expression by $\liminf$. We now introduce
\[
\tilde{\mu}_{n}\left(\mathcal{B},\mathbf{g}\right)=\underset{t\rightarrow\infty}{\liminf}\frac{1}{t}\sum_{l=0}^{t-1}\sum_{\tau=1}^{T}\mathbb{E}[u_{n,l}^{*}(\tau)|\mathcal{B},\mathbf{g}],
\]
and express 
\[
\mathbb{E}[u_{n,l-\tau}^{*}(\tau)]=\sum_{\mathbf{h}\in\mathcal{C}}\sum_{\mathcal{D}\subseteq\mathcal{N}}P_{c}(\mathbf{h})P_{d}(\mathcal{D})\mathbb{E}[u_{n,l-\tau}^{*}(\tau)|\mathbf{h},\mathcal{D}]
\]
 allowing us to write 
\begin{align}
c_{T}^{\mathcal{P}}(\bar{\boldsymbol{\pi}},\boldsymbol{\Psi}) & \geq\sum_{\mathbf{g}\in\mathcal{C}}\sum_{\mathcal{B}\subseteq\mathcal{N}}P_{c}(\mathbf{g})P_{d}(\mathcal{B})\,C_{d}\biggl(\Bigl[\delta_{n\in\mathcal{B}}\Bigl(S\nonumber \\
 & -\Bigl(\sum_{\mathbf{h}\in\mathcal{C}}P_{c}(\mathbf{h})\sum_{\mathcal{D}\subseteq\mathcal{N}}P_{d}(\mathcal{D})\,\tilde{\mu}_{n}\left(\mathcal{D},\mathcal{\mathbf{h}}\right)\Bigr)\Bigr)\nonumber \\
 & +\tilde{\mu}_{n}\left(\mathcal{B},\mathbf{g}\right)\Bigr]_{n=1}^{N};\mathbf{g}\biggr).
\end{align}
This proves the theorem. 

\section{Proof of Theorem \ref{th:asymptotic}}

\label{app:asymptotic} It suffices to prove that $\limsup_{T\to\infty}$
$c_{T}^{\wp_{\mathcal{U}}}(\bar{\boldsymbol{\pi}},\boldsymbol{\Psi})$
$=\liminf_{T\to\infty}$ $c_{T}^{\mathcal{P}}(\bar{\boldsymbol{\pi}},\boldsymbol{\Psi})$.
We start by $\limsup_{T\to\infty}c_{T}^{\wp_{\mathcal{U}}}(\bar{\boldsymbol{\pi}},\boldsymbol{\Psi})$.
Since policy $\wp_{\mathcal{U}}$ is stationary, we can ignore the
$\underset{t\rightarrow\infty}{\liminf}\frac{1}{t}\sum_{l=0}^{t-1}$
and write 
\begin{align*}
c_{T}^{\wp_{\mathcal{U}}}(\bar{\boldsymbol{\pi}},\boldsymbol{\Psi})= & \sum_{\mathcal{B}\subseteq\mathcal{N}}P_{d}(\mathcal{B})\sum_{\mathbf{g}\in\mathcal{C}}P_{c}(\mathbf{g})\\
 & \mathbb{E}\left[C_{d}\left(\Bigl[\delta_{n\in\mathcal{B}}(S-\sum_{\tau=1}^{T}u_{n,t-\tau}(\tau))\right.\right.\\
 & +\mu_{n}(\mathcal{B},\mathbf{g})\Bigr]_{n=1}^{N};\mathbf{g}\Biggr)|\mathcal{B},\mathbf{g}\Biggr],
\end{align*}
since $\sum_{\tau=1}^{T}u_{n,t}(\tau)=\sum_{\tau=1}^{T}T\mu_{n}(\mathcal{B},\mathbf{g})=\mu_{n}(\mathcal{B},\mathbf{g})$. 

Note that $\sum_{\tau=1}^{T}u_{n,t-\tau}(\tau)$ is independent of
$\mathcal{B}$, $\mathbf{g}$. We introduce a random variable $Z_{T}(\mathcal{D},\mathbf{h})$
which counts the number of occurrences of the pair of requesting set
$\mathcal{D}\subseteq\mathcal{N}$ and associated channel gain vector
$\mathbf{h}\in\mathcal{C}$, in slots $t-T$, $\cdots$, $t-1$. Then
\[
\sum_{\tau=1}^{T}u_{n,t-\tau}(\tau)=\sum_{D\subseteq\mathcal{N}}\sum_{\mathbf{h}\in\mathcal{C}}\frac{\mu_{n}(D,\mathbf{h})Z_{T}(D,\mathbf{h})}{T}
\]
By the strong law of large numbers, with probability 1, 
\[
\limsup_{T\to\infty}\frac{\mu_{n}(D,\mathbf{h})Z_{T}(D,\mathbf{h})}{T}=\mu_{n}(\mathcal{D},\mathbf{h})P_{d}(\mathcal{D})P_{c}(\mathbf{h})
\]
By noting that the system load at any time slot is uniformly bounded
above, bounded convergence theorem implies 
\begin{multline}
\limsup_{T\to\infty}c_{T}^{\wp_{\mathcal{U}}}(\bar{\boldsymbol{\pi}},\boldsymbol{\Psi})=\sum_{B\subseteq\mathcal{N}}P_{d}(\mathcal{B})\sum_{\mathbf{g}\in\mathcal{C}}P_{c}(\mathbf{g})C_{d}\left(\Bigl[\delta_{n\in\mathcal{B}}\left(S-\right.\right.\\
\left.\left.\sum_{\mathcal{D}\subseteq\mathcal{N}}\sum_{\mathbf{h}\in\mathcal{C}}\mu_{n}(\mathcal{D},\mathbf{h})P_{d}(\mathcal{D})P_{c}(\mathbf{h})\right)\right.\\
\left.+\mu_{n}(\mathcal{B},\mathbf{g})\Bigr]_{n=1}^{N};\mathbf{g}\right)=\underline{c}_{\mathcal{U}}(\bar{\boldsymbol{\pi}},\boldsymbol{\Psi}).
\end{multline}

Thus we have established that average expected cost under policy $\wp_{\mathcal{U}}$
attains the global lower bound as proactive service window size grows
to infinity. Now by the definition of $c_{T}^{\mathcal{P}}(\bar{\boldsymbol{\pi}},\boldsymbol{\Psi})$
being the minimum possible cost achieved by proactive scheduling with
proactive service window $T$, it follows that $\limsup_{T\to\infty}$
$c_{T}^{\wp_{\mathcal{U}}}(\bar{\boldsymbol{\pi}},\boldsymbol{\Psi})$
$=\liminf_{T\to\infty}$ $c_{T}^{\mathcal{P}}(\bar{\boldsymbol{\pi}},\boldsymbol{\Psi})$. 

\section{Proof of Theorem \ref{thm:TITV}\label{sec:Proof-of-Theorem_TITV}}

The optimal value is now 
\begin{equation}
c_{T}^{\mathcal{P}}(\bar{\boldsymbol{\pi}},\boldsymbol{\Psi}^{Q})=\underset{t\rightarrow\infty}{\limsup}\frac{1}{t}\sum_{l=0}^{t-1}\,\mathbb{E}\biggl[C_{d}\bigl(\left[L_{n,l}^{\mathcal{\mathcal{P}}}(\mathbf{u}_{n,l})\right]_{n=1}^{N};\mathbf{g}_{l}\bigr)\biggr].
\end{equation}
By joint conditioning on all possible sets of requesting users $\mathcal{B}_{l}$,
their possible channel state realizations $\mathbf{g}_{l}$, and the
slot indices $s$, we can write $c_{T}^{\mathcal{P}}(\bar{\boldsymbol{\pi},}\boldsymbol{\Psi})$
as 
\begin{align}
 & c_{T}^{\mathcal{P}}(\bar{\boldsymbol{\pi}},\boldsymbol{\Psi}^{Q})\nonumber \\
 & =\underset{t\rightarrow\infty}{\limsup}\frac{1}{t}\sum_{l=0}^{t-1}\sum_{\mathbf{g}\in\mathcal{C}}\sum_{s\in\mathcal{Q}}\sum_{\mathcal{B}\subseteq\mathcal{N}}\,P(\mathcal{B}_{l}=\mathcal{B},\mathbf{g}_{l}=\mathbf{g},s_{l}=s)\nonumber \\
 & \times\mathbb{E}\biggl[C_{d}\Bigl(\left[L_{n,l}^{\mathcal{\mathcal{P}}}(\mathbf{u}_{n,l})\right]_{n=1}^{N};\mathbf{g}_{l}\Bigr)\Bigr|\mathcal{B}_{l},\mathbf{g}_{l},s_{l}\biggr],
\end{align}
Clearly
\begin{align*}
 & P(\mathcal{B}_{l}=\mathcal{B},\mathbf{g}_{l}=\mathbf{g},s_{l}=s)\\
 & =P_{\mathrm{d}}(\mathcal{B}_{l}=\mathcal{B})P_{c}(\mathbf{g}_{l}=\mathbf{g}|s_{l}=s)P_{\mathrm{s}}(s_{l}=s)\\
 & =P_{\mathrm{d}}(\mathcal{B})P_{c}(\mathbf{g}|s)P_{\mathrm{s}}(s),
\end{align*}
due to the time-invariant nature of each distribution. Then
\begin{align}
 & c_{T}^{\mathcal{P}}(\bar{\boldsymbol{\pi}},\boldsymbol{\Psi}^{Q})\nonumber \\
 & =\sum_{\mathbf{g}\in\mathcal{C}}\sum_{s\in\mathcal{Q}}\sum_{\mathcal{B}\subseteq\mathcal{N}}P_{\mathrm{d}}(\mathcal{B})P_{c}(\mathbf{g}|s)P_{\mathrm{s}}(s)\nonumber \\
 & \times\underset{t\rightarrow\infty}{\limsup}\frac{1}{t}\sum_{l=0}^{t-1}\mathbb{E}\biggl[C_{d}\Bigl(\left[L_{n,l}^{\mathcal{\mathcal{P}}}(\mathbf{u}_{n,l}\right]_{n=1}^{N};\mathbf{g}_{l}\Bigr)\Bigr|\mathcal{B}_{l},\mathbf{g}_{l},s_{l}\biggr],
\end{align}
where the conditioning should be understood as $\mathcal{B}_{l}=\mathcal{B},\mathbf{g}_{l}=\mathbf{g},s_{l}=s$.
Now, substituting the definition of $L_{n,l}^{\mathcal{\mathcal{P}}}(\mathbf{u}_{n,l})$
from (\ref{eq:loadproactive}), we have 
\begin{align}
 & c_{T}^{\mathcal{P}}(\bar{\boldsymbol{\pi}},\boldsymbol{\Psi}^{Q})\nonumber \\
 & =\sum_{\mathbf{g}\in\mathcal{C}}\sum_{s\in\mathcal{Q}}\sum_{\mathcal{B}\subseteq\mathcal{N}}P_{\mathrm{d}}(\mathcal{B})P_{c}(\mathbf{g}|s)P_{\mathrm{s}}(s)\nonumber \\
 & \times\underset{t\rightarrow\infty}{\limsup}\frac{1}{t}\sum_{l=0}^{t-1}\mathbb{E}\biggl[C_{d}\Bigl(\Bigl[\delta_{n\in\mathcal{B}}\Bigl(S-\sum_{\tau=1}^{T}u_{n,l-\tau}(\tau)\Bigr)\nonumber \\
 & +\sum_{\tau=1}^{T}u_{n,l}(\tau)\Bigr]_{n=1}^{N};\mathbf{g}_{l}\Bigr)\Bigr|\mathcal{B}_{l},\mathbf{g}_{l},s_{l}\biggr].
\end{align}
Applying Jensen's inequality and accounting for term-wise conditional
independencies yields 
\begin{align}
 & c_{T}^{\mathcal{P}}(\bar{\boldsymbol{\pi}},\boldsymbol{\Psi}^{Q})\nonumber \\
 & \ge\sum_{\mathbf{g}\in\mathcal{C}}\sum_{s\in\mathcal{Q}}\sum_{\mathcal{B}\subseteq\mathcal{N}}P_{\mathrm{d}}(\mathcal{B})P_{c}(\mathbf{g}|s)P_{\mathrm{s}}(s)\nonumber \\
 & \times C_{d}\Bigl(\Bigl[\delta_{n\in\mathcal{B}}(S-\underset{t\rightarrow\infty}{\liminf}\frac{1}{t}\sum_{l=0}^{t-1}\sum_{\tau=1}^{T}\mathbb{E}[u_{n,l-\tau}(\tau)|s_{l}])\nonumber \\
 & +\underset{t\rightarrow\infty}{\limsup}\frac{1}{t}\sum_{l=0}^{t-1}\sum_{\tau=1}^{T}\mathbb{E}[u_{n,l}(\tau)|\mathcal{B}_{l},\mathbf{g}_{l},s_{l}]\Bigr]_{n=1}^{N};\mathbf{g}_{l}\Bigr).
\end{align}
We note that now $u_{n,l-\tau}(\tau)$ depends on the current channel
distribution (encoded through $s_{l}$), which is known at time $t-\tau$.
Similarly, $u_{n,l}(\tau)$ will depend on the future channel statistics,
at times $l+\tau$. Since $s_{l}$ is known, these statistics are
also known. Hence, we express
\begin{align*}
 & \sum_{\tau=1}^{T}\mathbb{E}[u_{n,l}(\tau)\Bigr|\mathcal{B},\mathbf{g},s]\\
 & =\sum_{\tau=1}^{T}\mathbb{E}[u_{n,l}(\tau)\Bigr|\mathcal{B},\mathbf{g},s_{l}=s,s_{l+\tau}=f_{\tau}(s)]
\end{align*}
in which $f_{\tau}(s)=\mod(s+\tau,Q)$, due to the cyclo-stationary
nature of the channel. We now define 
\begin{equation}
\underset{t\rightarrow\infty}{\limsup}\frac{1}{t}\sum_{l=0}^{t-1}\mathbb{E}[u_{n,l}(\tau)]\Bigr|\mathcal{B},\mathbf{g},s,s']=\frac{1}{T}\tilde{\mu}_{n}\left(\mathcal{B},\mathbf{g},s,s'\right),\label{eq:definitionmuHW}
\end{equation}
where $s,s'\in\{0,1,\ldots,Q-1\}$. Similarly, for $\underset{t\rightarrow\infty}{\liminf}\frac{1}{t}\sum_{l=0}^{t-1}\sum_{\tau=1}^{T}\mathbb{E}[u_{n,l-\tau}(\tau)|s_{l}]$,
we express as 
\begin{align*}
 & \underset{t\rightarrow\infty}{\liminf}\frac{1}{t}\sum_{l=0}^{t-1}\sum_{\tau=1}^{T}\mathbb{E}[u_{n,l-\tau}(\tau)|s_{l}]\\
 & =\frac{1}{T}\sum_{\mathcal{D}}P_{\mathrm{d}}(\mathcal{D})\sum_{\tau=1}^{T}\sum_{\mathbf{h}}P_{c}(\mathbf{h}|f_{-\tau}(s))\tilde{\mu}_{n}\left(\mathcal{D},\mathbf{h},f_{-\tau}(s),s\right).
\end{align*}
Finally, this leads to 
\begin{align}
 & c_{T}^{\mathcal{P}}(\bar{\boldsymbol{\pi}},\boldsymbol{\Psi}^{Q})\label{eq:generalcosttau}\\
 & \ge\sum_{\mathbf{g}\in\mathcal{C}}\sum_{s\in\mathcal{Q}}\sum_{\mathcal{B}\subseteq\mathcal{N}}P_{\mathrm{d}}(\mathcal{B})P_{c}(\mathbf{g}|s)P_{\mathrm{s}}(s)\nonumber \\
 & \times C_{d}\Bigl(\Bigl[\delta_{n\in\mathcal{B}}(S-\bar{\mu}_{n}(s))\nonumber \\
 & +\frac{1}{T}\sum_{\tau=1}^{T}\tilde{\mu}_{n}\left(\mathcal{B},\mathbf{g},s,f_{\tau}(s)\right)\Bigr]_{n=1}^{N};\mathbf{g}_{l}\Bigr).
\end{align}
with 
\[
\bar{\mu}_{n}(s)=\sum_{\mathcal{D}}P_{\mathrm{d}}(\mathcal{D})\frac{1}{T}\sum_{\tau=1}^{T}\sum_{\mathbf{h}}P_{c}(\mathbf{h}|f_{-\tau}(s))\tilde{\mu}_{n}\left(\mathcal{D},\mathbf{h},f_{-\tau}(s),s\right).
\]
The expression (\ref{eq:generalcosttau}) is the general version of
(\ref{eq:proactive_cost_TITV}), valid for any $T$. For the special
case of $T=LQ$, 
\begin{align*}
\bar{\mu}_{n}(s) & =\sum_{\mathcal{D}}P_{\mathrm{d}}(\mathcal{D})\frac{L}{T}\sum_{s'=0}^{Q-1}\sum_{\mathbf{h}}P_{c}(\mathbf{h}|s')\tilde{\mu}_{n}\left(\mathcal{D},\mathbf{h},s',s\right)\\
 & =\sum_{\mathcal{D}}P_{\mathrm{d}}(\mathcal{D})\sum_{s'=0}^{Q-1}P_{\mathrm{s}}(s')\sum_{\mathbf{h}}P_{c}(\mathbf{h}|s')\tilde{\mu}_{n}\left(\mathcal{D},\mathbf{h},s',s\right)
\end{align*}
and 
\[
\sum_{\tau=1}^{T}\tilde{\mu}_{n}\left(\mathcal{B},\mathbf{g},s,f_{\tau}(s)\right)=\sum_{s'=0}^{Q-1}P_{\mathrm{s}}(s')\tilde{\mu}_{n}\left(\mathcal{B},\mathbf{g},s,s'\right)
\]
in which case (\ref{eq:generalcosttau}) becomes (\ref{eq:proactive_cost_TITV}).
It should be noted that the definition (\ref{eq:definitionmuHW})
implies $0\leq\tilde{\mu}_{n}\left(\mathcal{B},\mathbf{g},s,s'\right)\leq S,\,\forall n,\mathcal{B},\mathbf{g},s,s'$.

\section*{Acknowledgment}

The authors would like to thank Suhail Ahmad, Rikard Reinhagen, and
Martin Dahlgren for their help in conducting the channel quality measurement
campaign.

\bibliographystyle{IEEEtran}
\phantomsection\addcontentsline{toc}{section}{\refname}\bibliography{Allerton_references}

\end{document}